\documentclass{article}
\usepackage[OT1]{fontenc} 
\usepackage[utf8]{inputenc}

\usepackage{spconf,amsmath,graphicx}
\usepackage{booktabs} 
\usepackage{amsmath}  
\usepackage{amsthm}
\usepackage{amsfonts}
\usepackage{array}
\newcolumntype{L}[1]{>{\raggedright\let\newline\\\arraybackslash\hspace{0pt}}m{#1}}
\newcolumntype{C}[1]{>{\centering\let\newline\\\arraybackslash\hspace{0pt}}m{#1}}
\newcolumntype{R}[1]{>{\raggedleft\let\newline\\\arraybackslash\hspace{0pt}}m{#1}}
\usepackage{caption}
\usepackage{lipsum}
\usepackage{xcolor}
\usepackage{wasysym}

\usepackage{floatrow}
\DeclareFloatFont{small}{\small}
\newcommand{\captiongray}[1]{\small{#1}}
\newcommand{\hyperfootnote}[1][]{\def\ArgI\hyperfootnoteRelay}
\newcommand\hyperfootnoteRelay[2][]{\href{#1#2}{\ArgI}\footnote{\href{#1#2}{#2}}}
\newcommand{\score}[2]{#1\textsuperscript{(#2)}}

\title{Cell Segmentation by Combining\\Marker-controlled Watershed and Deep Learning}
\name{Filip Lux, Petr Matula \thanks{The work was funded by the Czech Science Foundation, project no.~GA17-05048S. Access to computing and storage facilities owned by parties and projects contributing to the National Grid Infrastructure MetaCentrum provided under the program "Projects of Large Research, Development, and Innovations Infrastructures" (CESNET LM2015042), is greatly appreciated. We gratefully acknowledge the support of the NVIDIA Corporation and their donation of the Quadro P6000 GPU used for this research.}}
\address{Masaryk University, Faculty of Informatics\\
Centre for Biomedical Image Analysis\\
Botanick{$\acute{\text{a}}$} 68A, 602 00 Brno}

\begin{document}
%
\maketitle
\begin{abstract}
We propose a cell segmentation method for analyzing images of densely clustered cells.
The method combines the strengths of marker-controlled watershed transformation and a convolutional neural network (CNN).
We demonstrate the method universality and high performance on three Cell Tracking Challenge (CTC) datasets of clustered cells captured by different acquisition techniques.
For all tested datasets, our method reached the top performance in both cell detection and segmentation.
Based on a series of experiments, we observed:
(1) Predicting both watershed marker function and segmentation function significantly improves the accuracy of the segmentation.
(2) Both functions can be learned independently.
(3) Training data augmentation by scaling and rigid geometric transformations is superior to augmentation that involves elastic transformations.
Our method is simple to use, and it generalizes well for various data with state-of-the-art performance.

\end{abstract}

\begin{keywords}
Cell segmentation, Dense cell populations, Watershed, Convolutional neural networks
\end{keywords}

\section{INTRODUCTION}
\label{sec:intro}

Cell segmentation is a task of splitting a microscopic image domain into segments, which represent individual instances of cells.
It is a fundamental step in many biomedical studies, and it is regarded as a cornerstone of image-based cellular research.
Cellular morphology is an indicator of a physiological state of the cell \cite{Rittscher2010}, and a well-segmented image can capture biologically relevant morphological information \cite{Hill2007}.
Semi-automatic annotation requires many hours of manual curation, and it depends on procedures that are difficult to share between laboratories.
Because current automated image acquisition instruments produce large image datasets, fully automatic cell segmentation approaches are inevitable.

Automatic cell segmentation remains a challenging problem until today.
Although many approaches have already been developed \cite{Meijering2012, Xing2018}, for cells of more complex shapes or textures, automatic segmentation does not reach the quality of manual annotations \cite{CTC17}.
According to the results of the Cell Tracking Challenge (CTC), the research gap remains for datasets characterized by poor edge information, a low signal-to-noise ratio, low contrast ratio, and high cell density.

Marker-controlled watershed transform \cite{Meyer1990} is a mathematical morphology segmentation technique, which is historically popular for splitting clusters of biomedical objects.
According to the morphological segmentation paradigm \cite{Beucher1992}, watershed-based segmentation consists of selecting first a function indicating the object locations (marker function), and second a function quantifying the segmentation criterion to delineate object borders (segmentation function).
In cell segmentation, the watershed transform is commonly used for separating circular cells or cell nuclei \cite{Vaquero1998, Wahlby2004}.
However, with the increasing shape complexity of imagined objects, it becomes complicated to define both marker and segmentation functions properly.

In recent years, deep learning has become the methodology of choice in medical image processing \cite{Lecun2015, Litjens2017}.
The network called \hbox{u-net} significantly influenced the field of cell segmentation \cite{Ronneberger2015}.
It is a deep fully-convolutional neural network of hour-glass topology~\cite{Shelhamer2017}.
This model is supervised and needs to be adapted using a training dataset.
The \hbox{u-net} can segment cells of a complicated structure and shape, but it is not suitable for segmenting touching objects.

\begin{figure*}
  \includegraphics[width=1.00\textwidth]{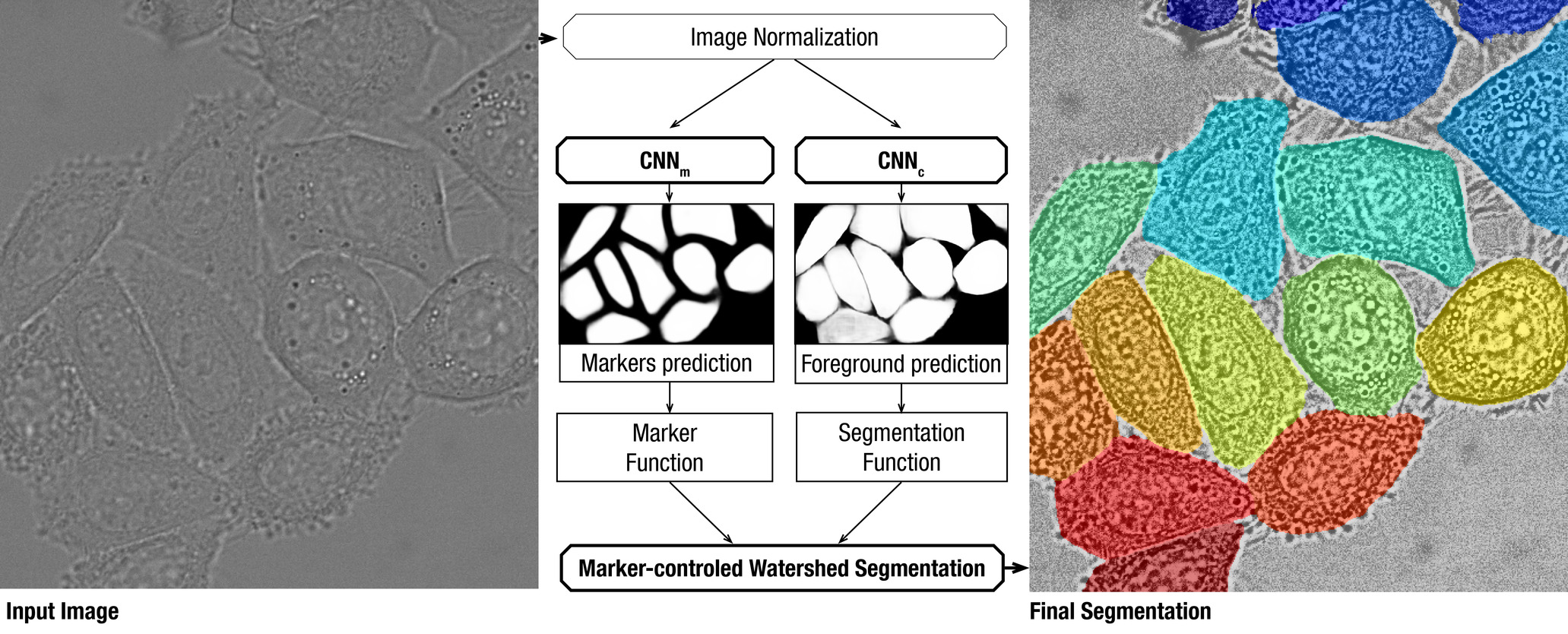}
  \caption{\textbf{Segmentation schema}: An input image is first normalized. A pair of CNNs predicts for each pixel probabilities of being a marker, and of being a cell mask. The CNN predictions are transformed into a marker function and a segmentation function using mathematical morphology operators. The final result is obtained by marker-controlled watershed segmentation.}
  \label{fig:schema}
\end{figure*}

To segment dense cell populations in difficult modalities, we propose to combine watershed transformation with deep learning.
We used two CNNs inspired by the topology of u-net that predict separately cell markers and image foreground (i.e., cell pixels).
From these predictions, we compute marker function and segmentation function of marker-controlled watershed segmentation (see Fig.~\ref{fig:schema}).
We pick three datasets of CTC that capture clustered cell populations, and that were obtained using different imaging techniques.
For these datasets, our approach reached the first or the second top score in CTC Cell Segmentation Benchmark (CSB), in terms of both segmentation and overall performance.

This paper is a significant extension of our preliminary work \cite{LuxISBI19}.
We intensively tested the method properties and improved the methodology and performance.
We verified its robustness for different image modalities by applying the method to two additional datasets, and we show that it can be used universally with top results.
The main article contributions are the following:
\begin{itemize}

\item We propose a method for segmenting dense cell populations in various modalities that improved the state-of-the-art segmentation accuracy in a public benchmark.
\item We introduce a pipeline based on mathematical morphology to extract both cell markers and image foreground from CNN predictions.
\item We identify and discuss critical components of the method that influence the performance.

\end{itemize}

\section{RELATED WORK}
\label{sec:related_work}


Cell segmentation is a wide-ranging topic, and many papers about segmentation approaches have been published over the years \cite{Meijering2012}. 
Recently with the increasing computational power of machines, deep learning \cite{Lecun2015} has become a technology of choice also in microscopy image analysis \cite{Litjens2017}.
In this field, the usage of CNNs \cite{LeCun1990} dominates, often as a part of a more complex image processing pipeline.

CNN is a type of neural network widely used for image processing, and it has been successful in many technological challenges \cite{Krizhevsky2017, Taigman2014, Ren2017}.
It consists of stacked convolutional layers that transform the input by the operation of convolution with a learnable kernel of a small size.
The depth of the network relates to the complexity of patterns that the network can recognize.

In microscopy image analysis, two main types of CNN topology are used nowadays.
The first one is based on the classification network of Krizhevsky et al. \cite{Krizhevsky2017}, which is used in various deep learning pipelines \cite{Chen2017, Li2018}.
The second type is based on fully convolutional networks of hour-glass topology called u-net \cite{Ronneberger2015}.
It focuses on local patterns of the image, and it extracts complex image features at different scales.
The hour-glass topology was shown to be successful for biomedical use, and it has become popular in many works.
It can be exploited for tasks of cell classification \cite{Zhang2017}, cell detection \cite{Sirinukunwattana2016, Xu2016}, and also cell segmentation \cite{Payer2019, Zhou2019}.
Our work follows the second branch of research. 


Image segmentation can be understood as a task of splitting an image domain into pairwise disjoint subsets, where each subset corresponds to a segmented object or the background.
In deep learning, the segmentation problem is commonly reformulated as a classification problem on a pixel level.
A straightforward way to segment cells is a binary classification of foreground and background pixels \cite{Ciresan2013, Ronneberger2015}, followed by labeling of the connected components of foreground pixels to get masks of individual cells.
A deep model learned in this way can successfully segment individual cells of irregular shape and appearance, but it fails for the segmentation of touching objects.

There are several ways how to segment clustered cells using CNN.
A frequent option is to define an additional class of pixels, which helps to split cell clusters.
This class can represent cell boundary \cite{VanValen2016, Chen2017}, pixels touching cells \cite{Guerrero-Pena2018}, cell nuclei \cite{Al-Kofahi2018}, cell centroids \cite{Zhou2019}, or cell membranes \cite{Eschweiler2019}.
The critical part is how to split the additional pixel class to get the final segmentation.
Some approaches \cite{VanValen2016, Chen2017} label detected cell boundaries as a background.
The others \cite{Guerrero-Pena2018, Al-Kofahi2018, Zhou2019} split touching cells based on various types of distances.
Such approaches work well for small convex cells, but are prone to errors if segmented cells have a complex shape.

Besides modifying the classification task, some approaches build their unique deep learning schema to segment clustered cells.
Yi et al.~\cite{Yi2019a} introduced an end-to-end deep learning pipeline to first detect cell instances by bounding boxes and then segment them individually.
The method is able to capture the slender and tiny structures of the cells.
Payer et al.~\cite{Payer2019} use a recurrent hour-glass network to predict pixel-wise embeddings of individual cell instances and use the mean shift clustering algorithm to assign the correct labels. In our work, we decided to use a different strategy, namely to employ the marker-controlled watershed transform. 

The marker-controlled watershed transformation \cite{Beucher1992} is a popular technique to segment clustered cells \cite{Vaquero1998, Lin2003, Pinidiyaarachchi2005, Yang2006, Wang2008, Koyuncu2012, Veta2013, Shu2013}.
All cited papers follow the morphological segmentation paradigm \cite{Beucher1992}, but they differ in the way how marker function and segmentation function are defined.
The definition usually utilizes knowledge of the structure and the shape of segmented cells.
Markers can be extracted by analyzing distance transform of a binarized image \cite{Vaquero1998, Lin2003}, by variance filtering refined by mathematical morphology operations \cite{Pinidiyaarachchi2005}, by image gradient analysis \cite{Wang2008, Veta2013}, or by utilizing morphological features~\cite{Koyuncu2012, Shu2013}.
The segmentation function is usually the input image or the image gradient.
This class of approaches segments correctly circular cells or cell nuclei with clearly visible borders, but their use is problematic for more complex input data.

The idea to combine the watershed transform with deep learning is not new~\cite{Al-Kofahi2018, Xie2020, Wang2019, Naylor2019}, but the way how it is combined in our approach and the related works is different.
The related works use a CNN only for object detection, i.e., for getting markers.
Al-Kofahi et al.~\cite{Al-Kofahi2018} predicts cell nuclei and use it as markers of watershed transformation.
Their approach was developed only for fluorescent images with stained cytoplasm.
Xie et al.~\cite{Xie2020} predicts cell markers and image foreground and uses the distance map of image foreground as a segmentation function.
The last two works~\cite{Wang2019, Naylor2019} use CNN to predict a Euclidian distance transform (EDT) of displayed objects.
Cell markers are then computed from a predicted EDT, either using mathematical morphology operators ~\cite{Naylor2019} or using a second CNN \cite{Wang2019}.
The predicted EDT is then used also as a watershed segmentation function.
The main drawback of these strategies is that they are inaccurate in the segmentation of cell boundaries.

\section{METHOD DESCRIPTION}
\label{sec:method_description}

The proposed method combines two prominent image processing techniques: a convolutional neural network and a marker-controlled watershed segmentation (see Fig.~\ref{fig:schema}).
The method is supervised.
It has to be initialized by training samples with reference annotations.
Before training, we augment training samples by scaling and by rigid geometric transformations.
We train two CNNs.
The first network predicts cell markers, and the second one predicts the image foreground (cell regions).
We apply a mathematical morphology pipeline to transform CNN predictions into a watershed \textit{marker function} and \textit{segmentation function}.
The final segmentation is produced by the marker-controlled watershed transformation.
In the following paragraphs, we describe each step of our method in detail.
\\

\subsection{Data preparation}
\label{ssec:preparation}
In biological and medical research, the usage of deep learning methods is difficult because of the lack of training data, and therefore various strategies, such as data normalization, data augmentation, weak annotations, and transfer learning, are often employed to make it possible \cite{Shorten2019}.
In our work, we normalize the input images to remove known biases in the datasets (Section \ref{sssec:normalization}), augment training samples to increase the dataset variability (Section \ref{sssec:augmentation_prep}), and use full as well as weak annotations (Section~\ref{sssec:annotations}) to prepare reference outputs for a successful network training (Section~\ref{sssec:markers}). 


\subsubsection{Data normalization}
\label{sssec:normalization}
The goal of data normalization is to unify the appearance of cell instances within the whole dataset while preserving the information relevant to the segmentation task.
For each input image $x$, we realize it by defining a normalization function $f_{\text{norm}}$.
The function is defined over the whole image domain $\Omega$, and it returns normalized image $x' = f_{\text{norm}}(x)$, whose values are always mapped to the range from $-0.5$ to $0.5$.

The optimal choice of the normalization function is data-dependent.
The source of the biases depends mainly on the image acquisition process, e.g., uneven illumination, level of illumination, or photo-bleaching.
As a normalization function, we tested histogram equalization (HE), contrast limited histogram equalization (CLAHE) \cite{Zuiderveld1994}, and median scaling, i.e., the image values were linearly scaled so that the overall median intensity was mapped to zero and the maximum value was mapped to 0.5.
The uneven image illumination can be removed, e.g., by high pass filtering or a top-hat transform.
We used different normalization functions for each of the tested datasets.
We describe our specific choice for each tested dataset in Sec.~\ref{ssec:datasets}.

\subsubsection{Data augmentation}
\label{sssec:augmentation_prep}

We use a data augmentation technique to introduce new patterns into the training dataset, which makes the training procedure more robust to over-fitting.
We generate new artificial samples from the original ones.
For the data augmentation, we use randomized rigid geometric transformations and scaling.
We rescale each training sample by the ratio from $0.6$ to $1.4$.
Then we rotate it by a random angle and eventually flip.
For all transformations, we extend data outside the domain by mirroring.
This augmentation strategy turned out to be the most appropriate based on our experiments described in Sec.~\ref{sssec:augmentation}.

\subsubsection{Training dataset annotations}
\label{sssec:annotations}
We consider two types of data annotations that we use for training: \textit{full annotation} and \textit{weak annotation}.
These annotations are typically created manually by experts.
A full annotation directly corresponds to an image segmentation.
It consists of \textit{cell masks} of individual cells, where each cell mask is a set of pixels that altogether represent a single cell.
A weak annotation is a union of cell markers that are compact subsets of cell masks (typically having a circular shape).
Each marker corresponds to one displayed cell.
Weak annotations are less time-consuming to create and cheaper to get than full annotations. 
Training datasets of CTC contain both types of annotations.

Our method requires a set of training samples with full annotations.
We can use weak annotations to train cell detection, which we discuss in the following section.
There is no required size of the training dataset.
In our experiments, we worked with networks trained on datasets of sizes from $4$ to $215$ training samples. In general, more training data implied better performance.

\subsubsection{Reference outputs}
\label{sssec:markers}

For neural network training, we use two types of reference outputs.
The first one is used to train cell detection by markers ($y_m$) and the second one to train the recognition of an image foreground ($y_c$).
Both outputs are binary images, and we compute them from available annotations.
The reference output $y_c$ includes all image pixels that represent any cell, and it is given by the full annotation.
The output $y_m$ is a union of cell markers, which are not touching to each other.
If we do not need to distinguish between $y_m$ and $y_c$, we use the symbol~$y$.

We can obtain cell markers either from weak or full annotations.
A weak annotation can be used directly as markers, only if they are not touching each other.
Otherwise, they have to be post-processed in the same way as full annotations.
To get markers from full annotations, we erode individual cell masks to introduce a gap in between them. 
The optimal way to define markers depends on the number and quality of available annotations.

We parametrized and studied the process of creating cell markers from full annotations.
To this end, we erode each cell mask by a circular structuring element with a diameter defined as:
$$d_{\text{SE}} = (1 - k) \cdot d_{max},$$
where $d_{max}$ is a diameter of the maximal disk included in the cell mask.
The parameter $k \in [0, 1]$ regulates the size of markers, where $1$ means no erosion and $0$ the ultimate erosion to a single point.
If the erosion results in a disconnected set of pixels, the largest connected component is taken as a marker.
In Sec.~\ref{sssec:msize}, we study the impact of parameter $k$ to the detection performance, and we compare it with a performance of markers defined by weak annotations.\\


\noindent
\begin{minipage}[b]{1.\linewidth}
  \centerline{\includegraphics[width=9.0cm]{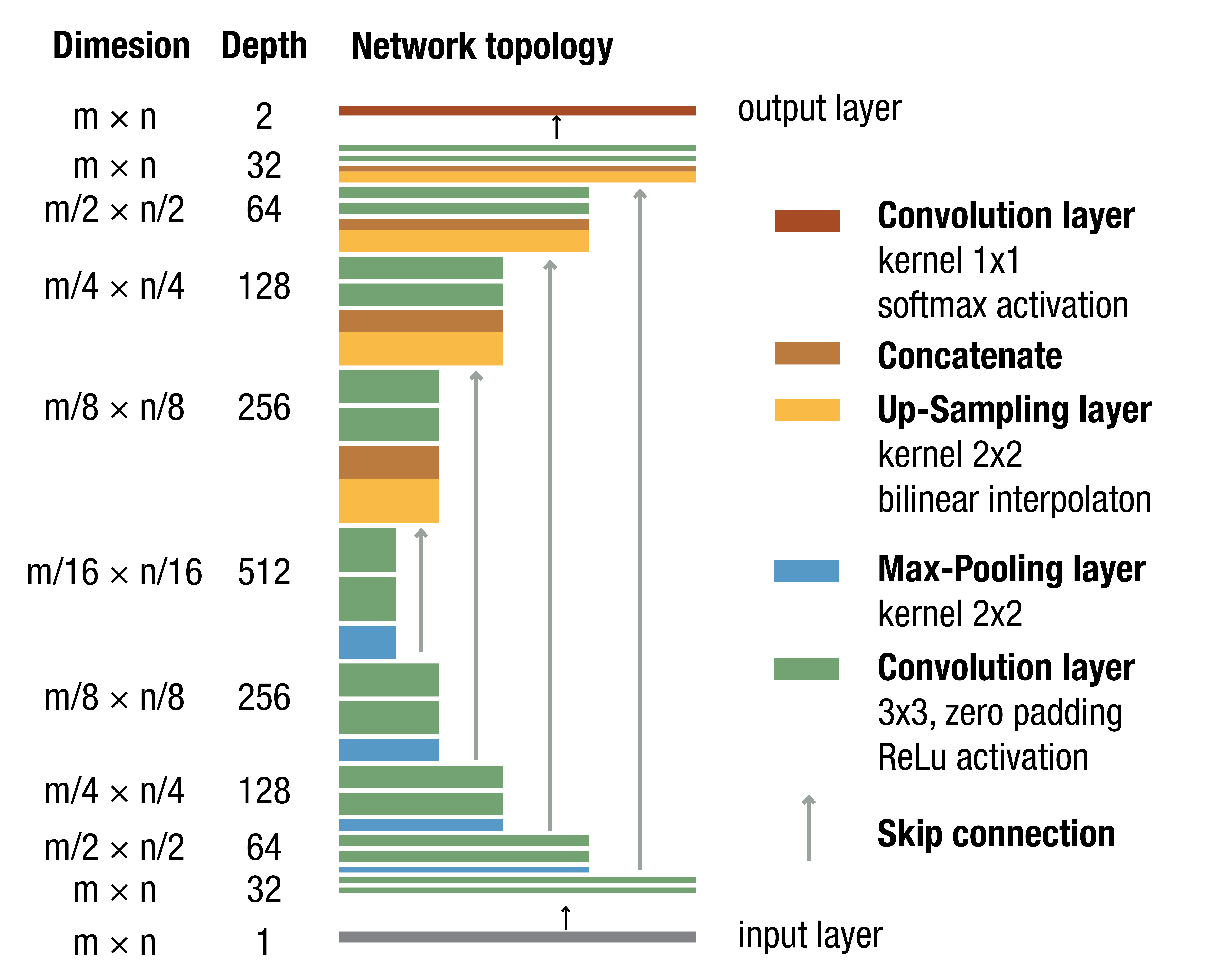}}
  \captionof{figure}{\captiongray{\textbf{The CNN architecture.} The topology is the same for both networks  $\text{CNN}_m$ and  $\text{CNN}_c$.}}
  \label{fig:network}
\end{minipage}

\subsection{CNN training}
\label{ssec:networks}

A cornerstone of our method is a pair of fully convolutional CNNs of hour-glass topology (see~Fig.~\ref{fig:network}).
We use one network ($\text{CNN}_m$) to predict cell marker pixels, and the second network ($\text{CNN}_c$) to predict the image foreground (cell regions).

We train these networks using the reference outputs $y_m$ and $y_c$, respectively.
Each reference output splits the pixels of the image domain into two classes $0$ and $1$.
The network predicts for each pixel probabilities $p_0(q)$, and $p_1(q)$ that the given pixel $q$ belongs to these classes.
The sum of these two probabilities is always $1$.
We map probabilities $p_1$ into the image domain, and we call the resulting image \textit{marker prediction} for network $\text{CNN}_m$, and \textit{foreground prediction} for network $\text{CNN}_c$.\\

\subsubsection{CNN architecture}
\label{sssec:nn_properties}

The architecture of our networks was inspired by a topology of the u-net~\cite{Ronneberger2015} with the following modifications:
Firstly, our network has fewer feature maps and learning parameters than u-net.
It makes our model more robust to over-fitting, especially when the training dataset contains a low number of samples.
It also lowers the computational time.
Secondly, for a convolution, we use zero-padding to produce an output image of the same size as the input image.
Thirdly, we use bilinear interpolation to up-sample feature maps in a decoder part. 
Lastly, we defined our own pixel weighting function incorporated into a loss function.

Each network consists of $18$ convolutional layers with kernel size $3 \times 3$.
The layers are formed into hour-glass topology with skip connections.
The last convolutional layer has a kernel of size $1 \times 1$ and a soft-max activation function.
We upsample feature maps using bilinear interpolation.
The input image dimensions $m, n$ have to be divisible by $16$.
The number of feature maps varies in the range from $2$ in an output layer to $512$ in the middle part of the network.


Because the network is fully convolutional, it has a valuable property that the number of learned parameters is the same for images of different sizes.
Therefore we can use the same model for any images of sizes $m \times n$ divisible by 16 without image resizing or tiling.
The divisibility by 16 is the only requirement for the correct behavior of max-pooling and up-sampling layers.
The images with other sizes can simply be padded with zeros or appropriately cropped to meet this requirement.

\subsubsection{Loss function}
\label{sssec:loss}

The training process is controlled by a loss function, which is a function that measures the error of a network prediction $p$ with respect to a reference output $y$.
We use weighted cross-entropy loss function that is computed as:
\begin{equation}\label{eq:wCE}
    L(p, y) = - \frac{\sum_{q \in \Omega} w(q)\log(p_{y(q)}(q))}{\sum_{q \in \Omega}  w(q) },
\end{equation}
where $w$ is a pixel weight function that is unique for every training sample.
To guarantee that every training sample has the same weight, we normalize the loss by the sum of all values in $w$.

The choice of our loss function was inspired by \cite{Ronneberger2015}, and we modified it for our application.
We generalized the pixel weight definition to prefer correct classification not only pixels in between cell masks, but all pixels close to the cell borders.
These pixels are essential both to divide cell markers and also to decide about the correct cell shape.
Note that we compared the performance of the pixel weight function from \cite{Ronneberger2015} with the weight function proposed in this paper, and we have found that our weights improve the results only slightly and the effect on the results presented in Section 4 is minor.
Nevertheless, we prefer our definition because it better meets the requirements of our application.

The definition of our pixel weight function is the following.
Let $\phi$ is a cell mask and $\Phi$ a set of masks of all the cells in the image.
To each pixel $q$ from the image domain $\Omega$ we assign a weight $w(q) \in \mathbb{R^+}$ by the formula:
\begin{equation}\label{eq:weights2}
    w(q) = [1 + a \sum\limits_{\phi \in \Phi } \max(d - ||q, \phi||, 0)] \cdot b,
\end{equation}
where $||q, \phi||$ is the Euclidian distance from $q$ to the closest pixel in $\phi$. 
By setting the parameter $a \in \mathbb{R^+}$, we regulate the weight magnitude.
We set it to $0.075$ to meet the range of u-net pixel weights.
The parameter $d \in \mathbb{R^+}$ relates to the width of the area around the object border with a higher weight.
The parameter $b$ balances the frequency of predicted classes.


\subsubsection{Training procedure}
\label{sssec:training}

We train both neural networks from scratch using the training schema that follows the standard practice and experience of similar approaches \cite{Payer2019, Isensee2019}.
We optimize the schema to our task by an analysis of learning curves and by a grid search.
We train each network for $12~800$ iterations, which are divided into $32$ epochs.
In one iteration, we process a mini-batch of $8$ randomly picked training samples.
Each sample is augmented on-the-fly by a randomized augmentation.
To find network parameters, we use Adam optimizer \cite{Kingma2014} with an initial learning rate of $3 \cdot 10^{-4}$.
During the training, we gradually decrease the learning rate down to the value of $3 \cdot 10^{-6}$.

\subsection{Marker function}
\label{ssec:marker_function}
The network $\text{CNN}_m$ predicts for each pixel a probability that it represents a marker.
We use this prediction to compute the watershed marker function.
The marker function is a binary image, where each connected component of pixels corresponds to a cell marker.
It is used in a marker-controlled segmentation to define segmentation seeds.

To get markers from the marker prediction image, we propose a marker extraction technique based on mathematical morphology.
The technique detects bright objects (cell markers) in marker prediction image and utilizes a prior knowledge about the marker shape.
Using this technique, we can control the quality of the extracted markers.
The technique has three parameters: marker diameter $d$, lower bound for a probability value $t_m$, and minimal contrast between neighboring markers~$h$.

The marker extraction has three-steps (see Fig.~\ref{fig:markers_postprocessing}).
First, we filter out all objects smaller than a minimal marker by the morphological opening.
The opening has a circular structuring element with a diameter $d$.
The parameter $d$ is given by equation $d = k \cdot d_{\text{inf}}$, where $d_{\text{inf}}$ is the minimum of maximal diameters of all cell masks in the training samples ($k$ in Sec. \ref{sssec:markers}).
Second, we remove structures with absolute values lower than $t_m$.
This step filters out objects with small probability to be a marker.
Third, we process the result by the h-dome (H-CONVEX) transform \cite{Soille1999} and pick all the pixels with a value equal to $h$ as marker pixels.
It ensures that the local contrast of each marker pixel measured by dynamics \cite{Grimaud92} is at least $h$.
Cell markers correspond to connected components of marker pixels.
In our experiments, we set $t_m$ to a constant value of $0.6$.\\

\noindent
\begin{minipage}[b]{1.\linewidth}
  \centerline{\includegraphics[width=8.5cm]{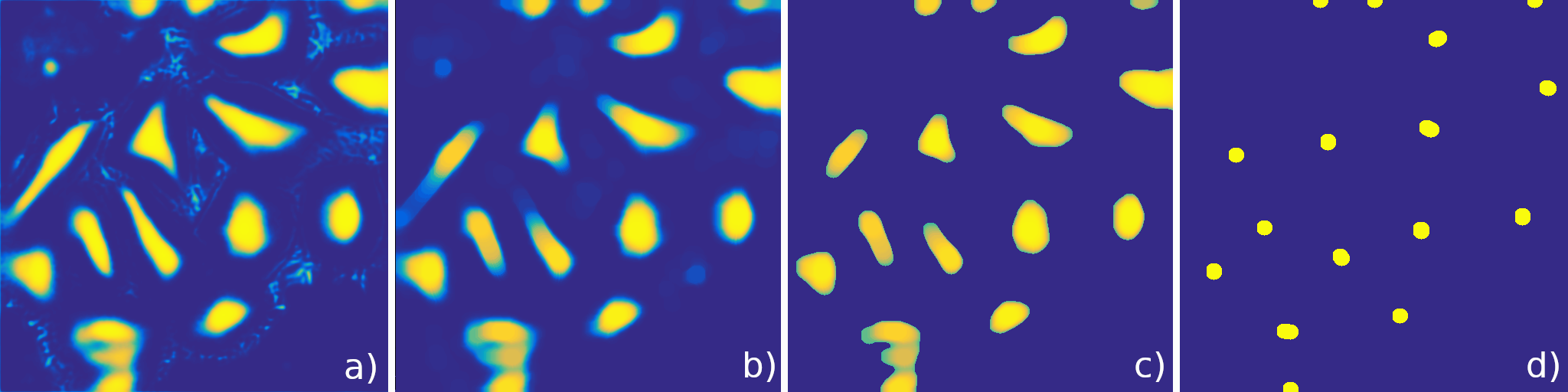}}
  \captionof{figure}{\captiongray{\textbf{Marker extraction}, (a) marker prediction image, (b) removing objects of small size, (c) removing objects of small probability to be a marker, (d) object splitting based on the contrast and final marker function} }
  \label{fig:markers_postprocessing}
\end{minipage}

\subsection{Segmentation function}
\label{ssec:seg_function}

We transform the foreground prediction calculated by the second network $\text{CNN}_c$  to the watershed \textit{segmentation function} and a mask of \textit{cell regions} as detailed here.

The desired segmentation function is an image, which has high values on object boundaries and low values in object interiors.
We compute it by inverting the foreground prediction image.
In this image, the probability values are small for pixels close to the cell boundary and high for pixels in the cell interior.\\

\noindent
\begin{minipage}[b]{1.\linewidth}
  \centerline{\includegraphics[width=8.5cm]{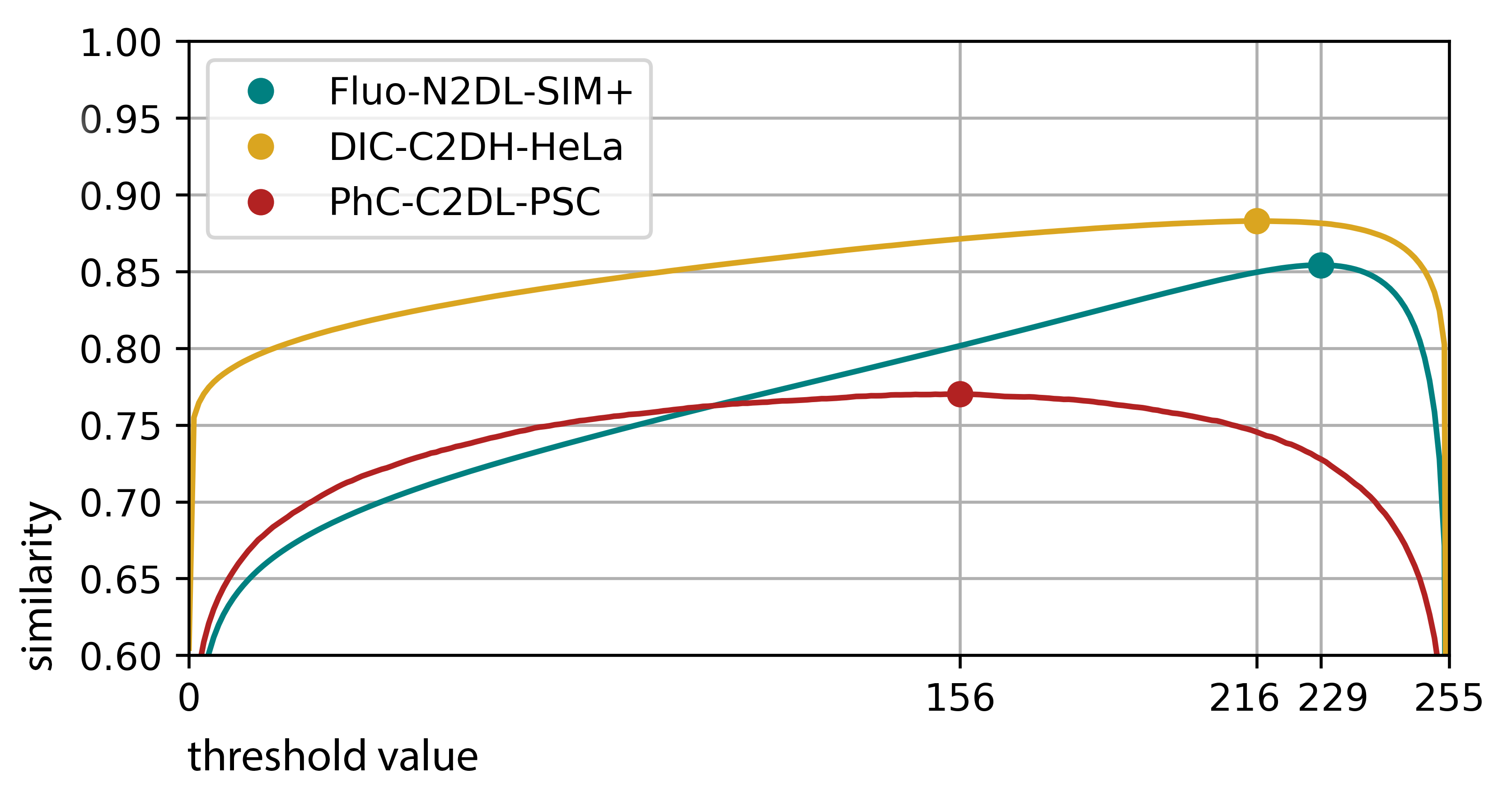}}
  \captionof{figure}{\textbf{Selection of the optimal threshold} to determine image background. We maximize the Jaccard similarity between thresholded foreground prediction image and a reference image.}
  \label{fig:tc_threshold}
\end{minipage}

To avoid the necessity of defining a background marker for the watershed transform, we calculate a mask of cell regions and run the watershed only in pixels representing cells.
The mask of cell regions is a set of pixels with a foreground prediction value higher than a threshold $t_c$.
The value $t_c$ is set to maximize the similarity between thresholded foreground prediction image and reference image $y_c$ over the training dataset.
We measure the similarity by the Jaccard similarity index.
In Fig.~\ref{fig:tc_threshold}, we show the relation between the threshold value and the similarity for three tested datasets. 
The optimal value of $t_c$ is data-dependent (see Sec \ref{ssec:datasets}), and to maximize performance, we set the threshold individually for each dataset.

\begin{figure*}
  \includegraphics[width=1.00\textwidth, height=0.40\textheight]{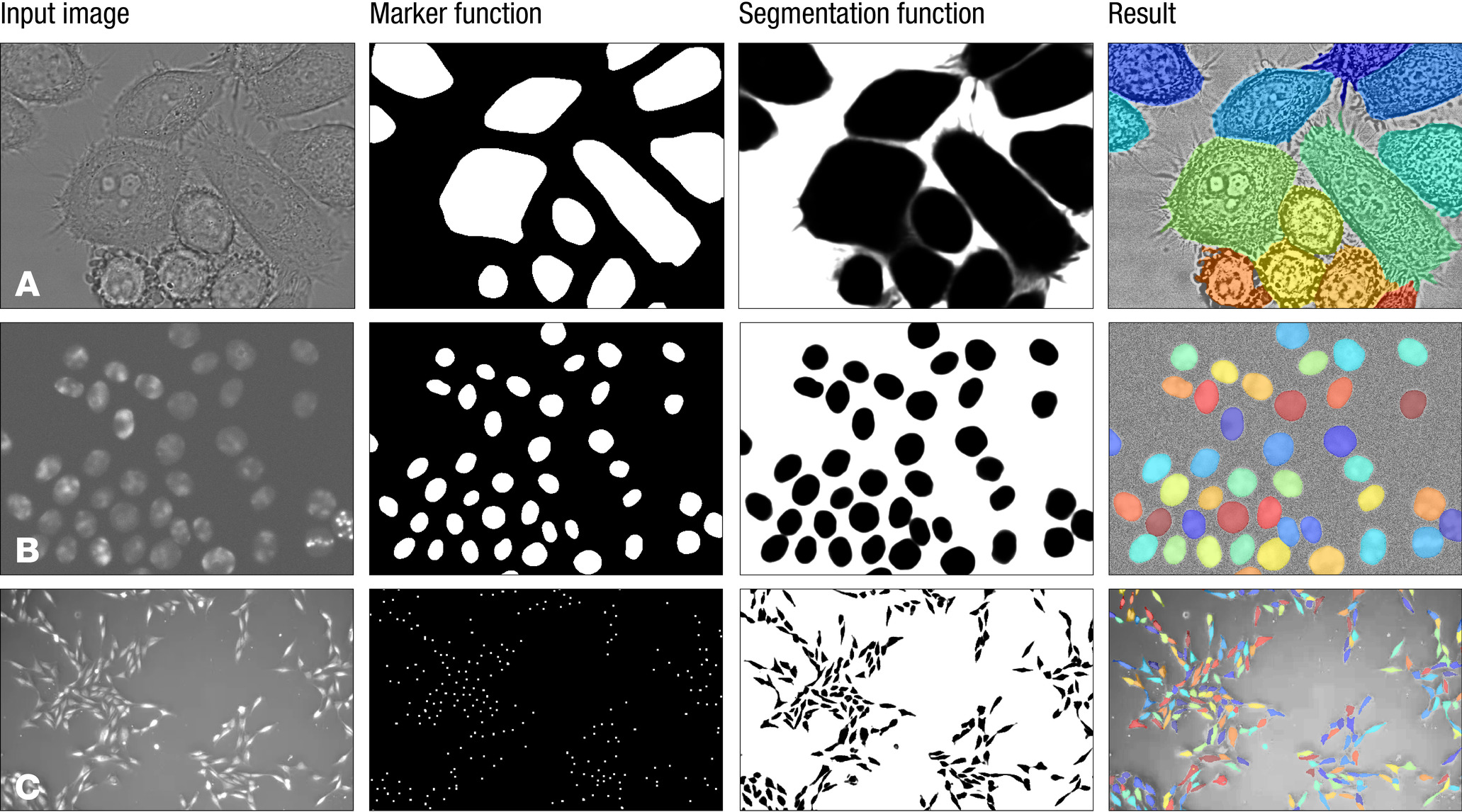}
  \caption{\textbf{Demostration of qualitative results} for three datasets from Cell Tracking Challenge: \textbf{A -- DIC-C2DH-HeLa} differential interference contrast, whole cells, HeLa cells; \textbf{B -- Fluo-N2DH-SIM+} fluorescence, simulated cell nuclei HL60; \textbf{C -- PhC-C2DL-PSC} phase contrast, whole cells, Pancreatic Stem Cells}
  \label{fig:results}
\end{figure*}

\subsection{Watershed segmentation}
\label{ssec:watershed}
Marker-controlled watershed segmentation is a non-parametric transformation of marker function, and segmentation function, which were defined in previous chapters.
Examples of both functions are shown in the middle columns of Fig.~\ref{fig:results}.
The watershed segmentation splits the image domain into segments.
In our work, we apply the watershed only to pixels within the cell regions.
The segmentation function controls the segmentation process.
The borderlines between the final segments follow the regions of high values of the segmentation function.
In the final segmentation, one segment corresponds to one marker.

If needed, we can refine the result by removing cell masks, which are touching the image boundary.
The cells are often excluded also from manual annotations.

\subsection{Technical details}
\label{ssec:technical_details}

Our implementation is written in Python3 using Keras library \cite{chollet2015keras} for deep learning and OpenCV library \cite{opencv_library} for image processing.
The training ran on a machine with a GPU NVIDIA Quadro P6000, and one epoch took approximately $210$ seconds.

\section{RESULTS}
\label{sec:results}

\begin{table*}[t]

  \centering
  \begin{tabular}{ L{2.2cm} | C{1.1cm} | C{1.1cm} | C{1.1cm} | C{0.cm} | C{1.1cm} | C{1.1cm} | C{1.1cm} | C{0.cm} | C{1.1cm} | C{1.1cm} | C{1.1cm} | l }
  \toprule
  \multicolumn{1}{c|}{} & \multicolumn{3}{c|}{DIC-C2DH-HeLa} & \multicolumn{1}{c|}{} & \multicolumn{3}{c|}{Fluo-N2DH-SIM+} & \multicolumn{1}{c|}{} & \multicolumn{3}{c|}{PhC-C2DL-PSC} &\\ 
  participant & SEG & DET & OP$_{\text{CSB}}$ & & SEG & DET & OP$_{\text{CSB}}$ & & SEG & DET & OP$_{\text{CSB}}$\\ 
  \midrule

  \textbf{MU-Lux-CZ} & \textbf{\score{0.863}{1}} & \textbf{\score{0.961}{1}} & \textbf{\score{0.912}{1}} & & \score{0.821}{2} & \score{0.971}{7} & \score{0.896}{2} & & \score{0.715}{2} & \score{0.967}{2} & \score{0.841}{2} &\\
  UVA-NL & \score{0.852}{2} & \score{0.958}{2} & \score{0.905}{2} & & \textbf{\score{0.822}{1}} & \score{0.972}{5} & \textbf{\score{0.897}{1}} & & \textbf{\score{0.720}{1}} & \textbf{\score{0.972}{1}} & \textbf{\score{0.846}{1}} &\\
  TUG-AT & \score{0.834}{3} & \score{0.956}{3} & \score{0.895}{3} & & \score{0.765}{11} & \score{0.979}{3} & \score{0.872}{9} & & - & - & - &\\
  BGU-IL\textsuperscript{(5)} & \score{0.820}{4} & \score{0.948}{4} & \score{0.884}{4} & & \score{0.790}{6} & \textbf{\score{0.983}{1}} & \score{0.887}{3} & & \score{0.654}{5} & \score{0.940}{7} & \score{0.797}{6} &\\
  FR-Ro-GE & \score{0.792}{7} & \score{0.902}{7} & \score{0.849}{7} & & \score{0.781}{8} & \score{0.981}{2} & \score{0.881}{6} & & \score{0.536}{14} & \score{0.902}{13} & \score{0.719}{13} &\\
  CVUT-CZ & \score{0.792}{5} & \score{0.906}{6} & \score{0.849}{6} & & \score{0.807}{3} & \score{0.959}{10} & \score{0.883}{5} & & \score{0.682}{3} & \score{0.936}{9} & \score{0.809}{3}&\\
  KTH-SE & \score{0.460}{10} & \score{0.855}{9} & \score{0.658}{9} & & \score{0.792}{5} & \score{0.960}{9} & \score{0.876}{7} & & \score{0.599}{11} & \score{0.966}{3} & \score{0.782}{9} &\\

  \bottomrule
  \end{tabular}
  \caption{\textbf{Cell Tracking Challenge - Cell Segmentation Benchmark results}, last update 12 Feb. 2020\\
  MU-Lux-CZ | Filip Lux, Masaryk University, Brno, Czech Republic; UVA-NL (based on MU-Lux-CZ)| Andreas Panteli, University of Amsterdam, Amsterdam, The Netherlands; TUG-AT | Christian Payer, Graz University of Technology, Graz, Austria; BGU-IL | Assaf Arbelle, Ben-Gurion University of the Negev, Beer-Sheva, Israel; FR-Ro-GE | Olaf Ronneberger, University of Freiburg, Freiburg, Germany; CVUT-CZ | Tomáš Sixta, Czech Technical University in Prague, Prague, Czech Republic; KTH-SE | Klas Magnusson, KTH Royal Institute of Technology, Stockholm, Sweden \& RaySearch Laboratories, Stockholm, Sweden}
  \label{tab:cell_tracking}
\end{table*}

\subsection{Datasets}
\label{ssec:datasets}

To demonstrate our method performance, we chose three CTC datasets: DIC-C2DH-HeLa, Fluo-N2DH-SIM+, and PhC-C2DL-PSC (see Fig.~\ref{fig:results} for examples).
These datasets display cells packed in clusters with low-contrast cell boundaries.
Datasets differ in cell size and cell appearance.
Two of them were captured by light microscopy techniques: differential interference contrast (\ref{fig:results}A) and phase-contrast microscopy (\ref{fig:results}C).
The third one consists of simulated fluorescence microscopy images (\ref{fig:results}B).

Each dataset of CTC consists of two training sequences released with reference annotations for method development and two challenge sequences used for an official evaluation by the challenge organizers.
Each training sequence includes full annotation of at least $200$ cell instances and weak annotation of the majority of the frames.
Full annotations were obtained by majority voting between three independent expert manual annotations.
The challenge sequences have strictly private reference annotations.

The first dataset of HeLa cells \textit{DIC-C2DH-HeLa} was captured in $2010$ by Gert van Cappellen from Erasmus Medical Center, Rotterdam, by differential interference contrast microscopy.
There are $84$ frames in each training sequence and $155$ frames in each testing sequence.
Only $16$ training images have full annotation. 
One frame is represented by a grayscale image of size $512 \times 512$.

The second dataset \textit{Fluo-N2DH-SIM+} contains simulated nuclei of HL60 cells stained with Hoescht, and it was synthesized by Vladimir Ulman and David Svoboda from Centre for Biomedical Image Analysis (CBIA), Masaryk University \cite{Svoboda_2017}.
In the training sequences, there are $65$ images of size $628 \times 690$ pixels, and $150$ images of size $739 \times 773$ pixels.
All training images have full annotation.

The third dataset of pancreatic stem cells \textit{PhC-C2DL-PSC} was captured by Dr. T. Becker from Fraunhofer Institution for Marine Biotechnology, Lübeck, Germany, in 2011 \cite{Rapoport2011}. 
In each training sequence, there are $426$ frames of size $720 \times 576$ pixels.
Only $4$ training images have full reference annotation. 
Because of the lack of full annotations, we trained $\text{CNN}_m$ using weak annotations.\\

\noindent
\begin{minipage}[b]{1.\linewidth}

   \center
  \begin{tabular}{ L{2.5cm} | C{1.5cm} | C{1.5cm} | C{1.5cm}l }
  
  \toprule
  dataset & A & B & C\\

  \midrule
  normalization & HE & HE & median\\
  markers -- type & eroded & eroded & weak\\
  $k$  & 0.8 & 0.8 & --\\
  $t_c$ -- threshold & 216 & 229 & 156\\
  $h$ -- dynamic & 5 & 30 & 3\\
  $d_{\text{min}}$ -- cell $\diameter$ (px) & 60 & 20 & 6\\

  \bottomrule
  \end{tabular}
  \captionof{table}{Dataset specific parameters}
 
  \label{tab:parameters}

\end{minipage}

\subsection{Evaluation metrics}
\label{ssec:evaluation}
We evaluate the method from the perspective of cell segmentation and cell detection by SEG measure and DET measure, which are the official measures of CTC Cell Segmentation Benchmark (CSB).
The measures compare the method segmentation and detection results with respect to reference annotations.
Their values are between $0$ and $1$, and a higher value corresponds to better performance. 
In the benchmark, the methods are compared by the arithmetic mean of both measures called overall performance (OP$_{\text{CSB}}$).

The SEG measure is to understand how well the segmented regions match the reference regions.
Let $R$ be a set of pixels of a reference region, and $S$ be a set of pixels of one segmented region.
The set $S$ matches to the set $R$ if it holds $|R \cap S| > 0.5 \cdot |R|$.
The Jaccard similarity index then measures the segmentation score $J$ in between regions $R$ and~$S$:
\begin{equation}\label{eq:SEG}
    J(R, S) = \frac{|R \cap S|}{|R \cup S|}.
\end{equation}
If there is no matching region $S$, then the score is zero. The SEG measure is a mean of $J$ indices of all regions $R$.

The DET measure is computed by normalized Acyclic Oriented Graph Matching measure for detection (AOGM-D) \cite{Matula_2015}.
The matching condition between a reference object and a segment is the same as for the SEG measure.
Perfect matching has DET measure equal to $1$.
The measure penalizes the following three events: (1) the reference object does not match any segment, (2) the cell segment matches no reference object, (3) and the segment matches more than one reference object.
These events are weighted by constants 10, 1, and 5, respectively, as suggested in \cite{Matula_2015}.
The measure is normalized by the total number of reference objects, and the measure can not be less than $0$.

\subsection{Evaluation}
\label{ssec:evaluation}
To get objective quantitative results, we submitted our method to the CSB benchmark.
This benchmark is open for new submissions over the year, and the evaluation is made by challenge organizers on private testing data.
At the time of our submission in September 2019, our method achieved the first overall score for all tested datasets.
The method excelled especially in SEG measure, where it reduced the segmentation error relatively by up to $17\%$.
The updated results from February 20th, 2020, are listed in Table~\ref{tab:cell_tracking}.
In this snapshot, our method was overcome by Andreas Panteli et al. from the University of Amsterdam (UVA-NL), who directly used our results and improved accuracy by modeling of temporal events through the sequences.
All competing methods are described in detail on the challenge website\footnote{https://www.celltrackingchallenge.org/participants/}.

A demonstration of qualitative results for tested datasets is in Fig.~\ref{fig:results}. 
Our method worked well for all tested data.
The size of segmented cells varied, and the method also worked for cells of a noncircular shape.
Because of the controlled cell detection by markers, the method did not suffer from over-segmentation or merging of adjacent cells.
The shape of the segmented cells was visually correct.

\subsection{Experiments}
\label{ssec:experiments}
To provide a deeper view of the method behavior, we report the results of four experiments.
In these experiments, we study: (1) the effect of three data augmentation methods; (2) the impact of predicted segmentation function to the performance (3); different definitions of cell markers; (4) and two different deep learning schemes.

All the experiments were carried out primarily on the DIC-C2DH-HeLa dataset.
Because the challenge dataset annotations are available only for CTC organizers, we use CTC training sequences for both training and validation.
For the purpose of these experiments, we manually annotated both training sequences to be able to present more objective results.
The subsequent frames in one sequence are correlated, therefore we validated the method across the sequences.
The experiments were evaluated by SEG and DET measures, where SEG measure corresponds directly to the quality of foreground prediction, and the DET measure relates to the quality of marker function.
In every experiment, we measured the model performance in 20 time-steps at the end of the training procedure, and then we reported the mean performance score.

\begin{figure*}
  \includegraphics[width=1.00\textwidth]{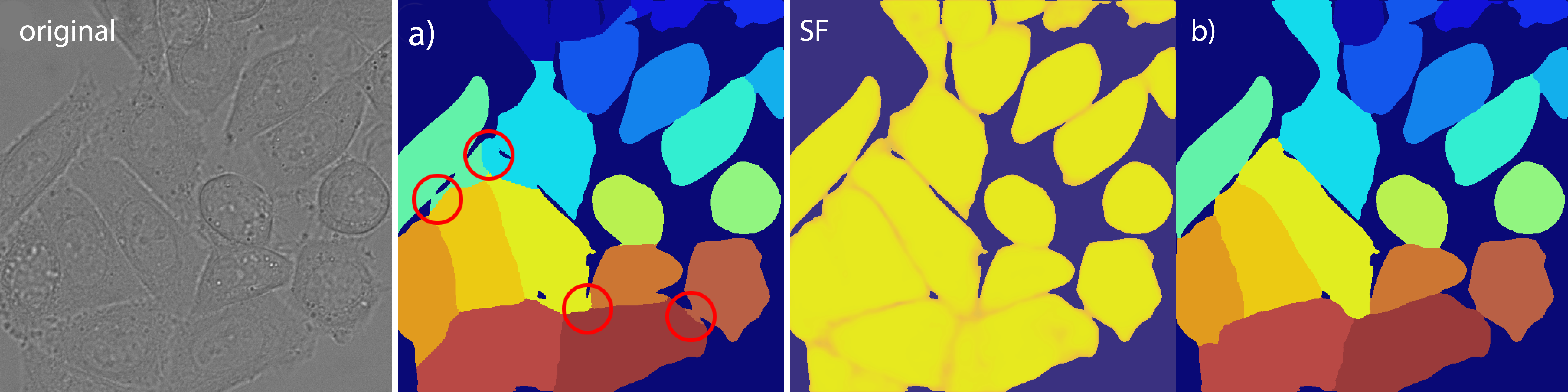}
  \caption{\textbf{Comparison of our result with related approaches} for DIC-C2DH-HeLa dataset. In our approach, we predict both marker function and segmentation function; (original) -- Original image; (a) -- Result of a watershed controlled by segmentation function based on the distance. Errors are marked by red circles; (SF) -- Predicted segmentation function; (b) -- Our result using the predicted segmentation function.}
  \label{fig:watershed}
\end{figure*}

\subsubsection{Data augmentation}
\label{sssec:augmentation}
The first experiment compares the efficiency of data augmentation by geometric transformations with or without elastic deformations.
We tested image augmentation by rigid geometrical transformations and scaling (RTS), and augmentation by elastic deformations (ED) \cite{Simard2003}.
As a baseline, we trained the network on samples without any data augmentation (no AUG).
It turned out that it is best to use only rigid transformations and scaling (RTS) for data augmentation (Tab.~\ref{tab:augmentation}).
The usage of ED is better than no augmentation, but our experiment indicates that it is not beneficial to combine it with RTSs.
Elastic deformations enrich the training dataset by new patterns, but it also introduces a high amount of noise.
Following this observation, in our work, we augmented data only by rigid deformations and scaling.\\
\\
\noindent
\begin{minipage}[b]{1.\linewidth}
\small{
   \center
  \begin{tabular}{ L{1.6cm} | C{1.2cm} | C{1.2cm} | C{1.2cm} | C{1.2cm}l   }
  
  \toprule
  dataset & no AUG & ED & ED+RTS & RTS\\

  \midrule
  DIC -- SEG & 0.749 & 0.787 & 0.844 & \textbf{0.851} \\
  DIC -- DET & 0.956 & 0.966 & 0.969 & \textbf{0.973}\\

 \bottomrule
  \end{tabular}
  \captionof{table}{\textbf{Augmentation techniques} and their impact on the accuracy of detection and segmentation. no AUG -- no data augmentation, ED -- elastic transformations, RT -- rigid geometric transformations and scaling }
 
  \label{tab:augmentation}
}
\end{minipage}

\subsubsection{Segmentation function}
\label{sssec:segfunction}
In the second experiment, we studied whether the predicted segmentation function has a positive effect on the segmentation quality (Fig.~\ref{fig:watershed}).
As a baseline, we assign labels to foreground pixels based on a distance to the closest marker (a).
It corresponds to the definition of segmentation function in~\cite{Xie2020} or the fine segmentation procedure in~\cite{Zhou2019}.
We compared it with our approach (b), where the segmentation process is driven by the predicted segmentation function.
We observed that the distance to a cell marker is not a sufficient clue to label the image foreground accurately.
The quality of (a) depends on the cell marker size and location.
It is low if cells are touching each other, and markers are not in the cell center.
The result of (b) better segments touching cells, and it is independent of marker location and size.

Quantitative comparison is presented in Tab.~\ref{tab:seg_function}.
The usage of our predicted segmentation function increased the accuracy of segmentation by approximately $3$ percentage points.
The improvement is larger if there are clusters of touching cells.\\

\noindent
\begin{minipage}[b]{1.\linewidth}
\small{

   \center
  \begin{tabular}{ L{3.8cm} | C{1.0cm} | C{1.0cm} | C{1.0cm}l }
  
  \toprule
  dataset  & (a) & (b) & diff\\

  \midrule
  DIC -- SEG & 0.795 & \textbf{0.828} & +~0.033 \\
  PhC -- SEG & 0.677 & \textbf{0.691} & +~0.024 \\

 \bottomrule
  \end{tabular}
  \captionof{table}{\textbf{Difference between two segmentation functions (SFs)} The performance for distance based SF (a); our predicted SF (b); and the difference between the two (diff).}
 
  \label{tab:seg_function}
}
\end{minipage}

\subsubsection{Marker type}
\label{sssec:msize}
In the third experiment, we compared the detection performance of markers defined from full annotations and weak annotations (Tab.~\ref{tab:markers}).
We also wanted to understand the impact of the parameter $k$ that relates to the marker size.

We observed markers defined from full annotations were more suitable for cell detection than markers from weak annotations.
We suppose that it is because the markers in weak annotations do not always lie in the cell center.
The marker location is then ambiguous, which can lead to detection errors.
This error is significant for larger cells with complex shapes.
Parameter $k$ does not have a significant impact on detection performance.
Generally, it can be set in the interval from $0.4$ to $0.6$ with similar results.\\
%

\noindent
\begin{minipage}[b]{1.\linewidth}
\small{

   \center
  \begin{tabular}{ L{1.5cm} | C{0.8cm} | C{0.8cm} | C{0.8cm} | C{0.8cm} | C{1.3cm}l }
  
  \toprule
    & \multicolumn{4}{c|}{full annotations} & weak ann. \\
  dataset \textbackslash ~ k & 0.2 & 0.4 & 0.6 & 0.8 & -  \\

  \midrule
  DIC -- DET & 0.971 & 0.979 & 0.976 & 0.978 & 0.954\\
    
  \bottomrule
  \end{tabular}
  \captionof{table}{\textbf{Detection performance} for trained markers. The evaluation measure is the DET measure. $k$ -- a ratio of cell diameter and a marker diameter, weak -- markers defined by weak annotations}
 
  \label{tab:markers}
}
\end{minipage}

\subsubsection{Multi-task learning}
\label{sssec:multitask}
In the last experiment, we compared the performance of two separate networks for marker prediction and foreground prediction with the performance of only one network predicting both tasks at once, as was used in \cite{LuxISBI19}.
The motivation to train only one network was to improve network stability by learning more general features.
We observed that both options have a similar performance in detection and also in segmentation.

We observed that multi-task learning has no significant positive effect.
The difference is not higher than 0.003 in both measures.
In our method, we decided to use two networks.
The main advantage is that it simplifies handling with various sources of training annotations for marker and foreground prediction, and different networks can be easily combined.\\

\noindent
\begin{minipage}[b]{1.\linewidth}
\small{

   \center
  \begin{tabular}{ L{2.5cm} | C{1.9cm} | C{1.9cm}l }
  
  \toprule
  dataset & multi-task & this paper \\

  \midrule
  DIC -- SEG & 0.826 & 0.823\\
  DIC -- DET & 0.923 & 0.926\\
    
  \bottomrule
  \end{tabular}
  \captionof{table}{\textbf{Multi-task learning} We tested the performance gain by using a multi-task learning technique:  One network trained for both marker prediction and foreground prediction (multi-task); two separated networks (this paper).}
 
  \label{tab:multi_task}
}
\end{minipage}

\section{CONCLUSION AND FUTURE WORK}
\label{sec:conclusion}
We described a cell segmentation method suitable for the segmentation of clustered cells, which achieved state-of-the-art performance for three different datasets in terms of a public cell segmentation benchmark.
The method is based on two CNNs, the first to detect cell markers, and the second to predict the image foreground.
We propose a process to transform these predictions into a watershed marker function and a segmentation function.
The final segmentation is then produced directly by a marker-controlled watershed transform.
We tested the method on three qualitatively different datasets of dense cell populations from the Cell Tracking Challenge (CTC).
For all the datasets, the method reached the top scores in cell segmentation as well as cell detection.
The proposed method showed to be a viable technique for the segmentation of dense cell populations.


Based on a series of practical experiments we have found that
(1) using only rigid transformations and scaling to augment training datasets is superior to also incorporate elastic transformations,
(2) predicted watershed segmentation function improves the watershed segmentation accuracy compared to related works,
(3) learning markers from eroded full annotations is superior to using only weak annotations, where the erosion size does not have a dramatic influence on the obtained results.

In future work, we intend to generalize the method for volumetric data and to utilize also the temporal information.

\bibliographystyle{IEEEbib}
\setlength{\parindent}{1em}
\bibliography{library}

\end{document}